\documentclass[twocolumn, amsmath, superscriptaddress]{revtex4-2}

\usepackage{graphicx}
\usepackage{dcolumn}
\usepackage{braket}
\usepackage{bm}
\usepackage[breaklinks=true,colorlinks,citecolor=blue,linkcolor=blue,urlcolor=blue]{hyperref}
\usepackage[dvipsnames]{xcolor}
\usepackage{subfigure}
\usepackage{tabularx}
\usepackage{amssymb}
\usepackage{gensymb}
\usepackage{multirow}
\usepackage[normalem]{ulem}

\newcolumntype{L}[1]{>{\raggedright\let\newline\\\arraybackslash\hspace{0pt}}m{#1}}
\newcolumntype{C}[1]{>{\centering\let\newline\\\arraybackslash\hspace{0pt}}m{#1}}
\newcolumntype{R}[1]{>{\raggedleft\let\newline\\\arraybackslash\hspace{0pt}}m{#1}}
\newcommand{\Ha}{\mathcal{H}}

\begin{document}
\title{Convolutional transformer wave functions}

\author{Ao Chen}
\altaffiliation{These authors contributed equally to this work}
\affiliation{Center for Electronic Correlations and Magnetism, University of Augsburg, 86135 Augsburg, Germany}
\affiliation{Center for Computational Quantum Physics, Flatiron Institute, New York 10010, USA}

\author{Vighnesh Dattatraya Naik}
\altaffiliation{These authors contributed equally to this work}
\affiliation{Center for Electronic Correlations and Magnetism, University of Augsburg, 86135 Augsburg, Germany}

\author{Markus Heyl}
\affiliation{Center for Electronic Correlations and Magnetism, University of Augsburg, 86135 Augsburg, Germany}

\def\thefootnote{*}\footnotetext{These authors contributed equally to this work}\def\thefootnote{\arabic{footnote}}

\begin{abstract}
Deep neural quantum states have recently achieved remarkable performance in solving challenging quantum many-body problems.
While transformer networks appear particularly promising due to their success in computer science, we show that previously reported transformer wave functions haven't so far been capable to utilize their full power.
Here, we introduce the convolutional transformer wave function (CTWF).
We show that our CTWFs exhibit superior performance in ground-state search and non-equilibrium dynamics compared to previous results, demonstrating promising capacity in complex quantum problems.

%
%
\end{abstract}

\maketitle

\textit{Introduction.}---
The accurate numerical solution of strongly correlated quantum matter remains as an outstanding challenge in modern quantum physics.
This concerns in particular the regime of large two-dimensional quantum many-body systems despite of impressive theoretical developments.
In recent years the neural quantum state (NQS) has emerged as a promising numerical method to solve the quantum many-body problem.
The NQS is based on utilizing artificial neural networks (ANNs) to encode the quantum many-body wave functions~\cite{Carleo_Science17_NQS}.
As ANNs are universal function approximators the NQS becomes a numerically exact approach converging, in principle, to the exact solution upon increasing the size of the ANN.
Until now, the NQS has shown great potential in various quantum many-body problems, including quantum spin liquids~\cite{Nomura_PRX21_PPRBMJ1J2, Astrakhantsev_PRX21_pyrochlore, Chen_NP24_MinSR}, Fermi-Hubbard models~\cite{Nomura_PRB17_PPRBM, Luo_PRL19_Backflow, Moreno_PNAS22_FermiNQSconstrain}, electronic structures~\cite{Herman_NC23_NQSchem, Choo_NC20_QuChem, Pfau_PRR20_FermiNet, Hermann_NC20_PauliNet, Pfau_Science24_ChemExcited}, open quantum systems~\cite{Alexandra_PRL19_OpenNQS, Hartmann_PRL19_OpenNQS, Filippo_PRL19_OpenNQS}, and quantum dynamics~\cite{Schmitt_PRL20_NQSdynamics, Schmitt_SA22_QPTdynamics, Sinibaldi_Q23_UnbiaseTDVP, Mendes_2023, Nys_arxiv24_ElectronDynamics, Mendes_2024}.

The key enabling potential for the NQS technique is the expressive power of the underlying ANNs, in particular when it comes to modern deep architectures.
While the initial starting points were based on still relatively shallow networks such as restricted Boltzmann machines (RBMs)~\cite{Carleo_Science17_NQS, Nomura_JPCM2021_RBMsymm}, in the recent years many deeper networks have already been investigated, including convolutional neural networks (CNNs)~\cite{Choo_PRB19_J1J2CNN}, variational autoregressive networks~\cite{Sharir_PRL20_QVAN}, recurrent neural networks (RNN)~\cite{Hibat-Allah_PRR20_NQSRNN}, group CNNs~\cite{Roth_PRB23_GCNN}, and deep CNNs~\cite{Liang_MLST23_CNNJ1J2, Chen_NP24_MinSR}. As a consequence of recent advances in training algorithms, it has now also become possible to optimize deep NQSs with up to $10^6$ parameters thereby pushing the NQS approach more towards exploiting the full power of ANNs. In particular, with such deep NQSs unprecedented numerical accuracies have been reached for multiple frustrated quantum magnets~\cite{Chen_NP24_MinSR}.

In the field of machine learning, transformer neural networks have developed into the most powerful architecture for many tasks~\cite{Vaswani_NIPS17_attention, Dosovitskiy_ICLR21_ViT}.
Consequently, it is a natural immediate question to which extent such transformer architectures could have a similar powerful potential for NQS.
So far, in the context of NQS transformers have already been applied in electronic structure problems~\cite{vonGlehn_arxiv23_Psiformer, Pescia_PRB24_MessagePassing, Shang_arxiv24_QiankunNet, Pfau_Science24_ChemExcited} and quantum lattice models~\cite{Viteritti_PRL23_ViT, Zhang_PRB23_MaskTransformer}. However, as we will discuss in this work, these previously introduced design choices in particular for lattice models are not in such a form yet so as to fully exploit the power of transformer architectures: they either are actually equivalent to CNNs, and therefore don't fully exploit the key attention mechanism in transformers, or break translation symmetries. Thus, it has so far remained open how to fully utilize the full potential of transformer architectures in the context of quantum lattice models.

\begin{figure*}[t]
    \centering
    \includegraphics[width=0.9\linewidth]{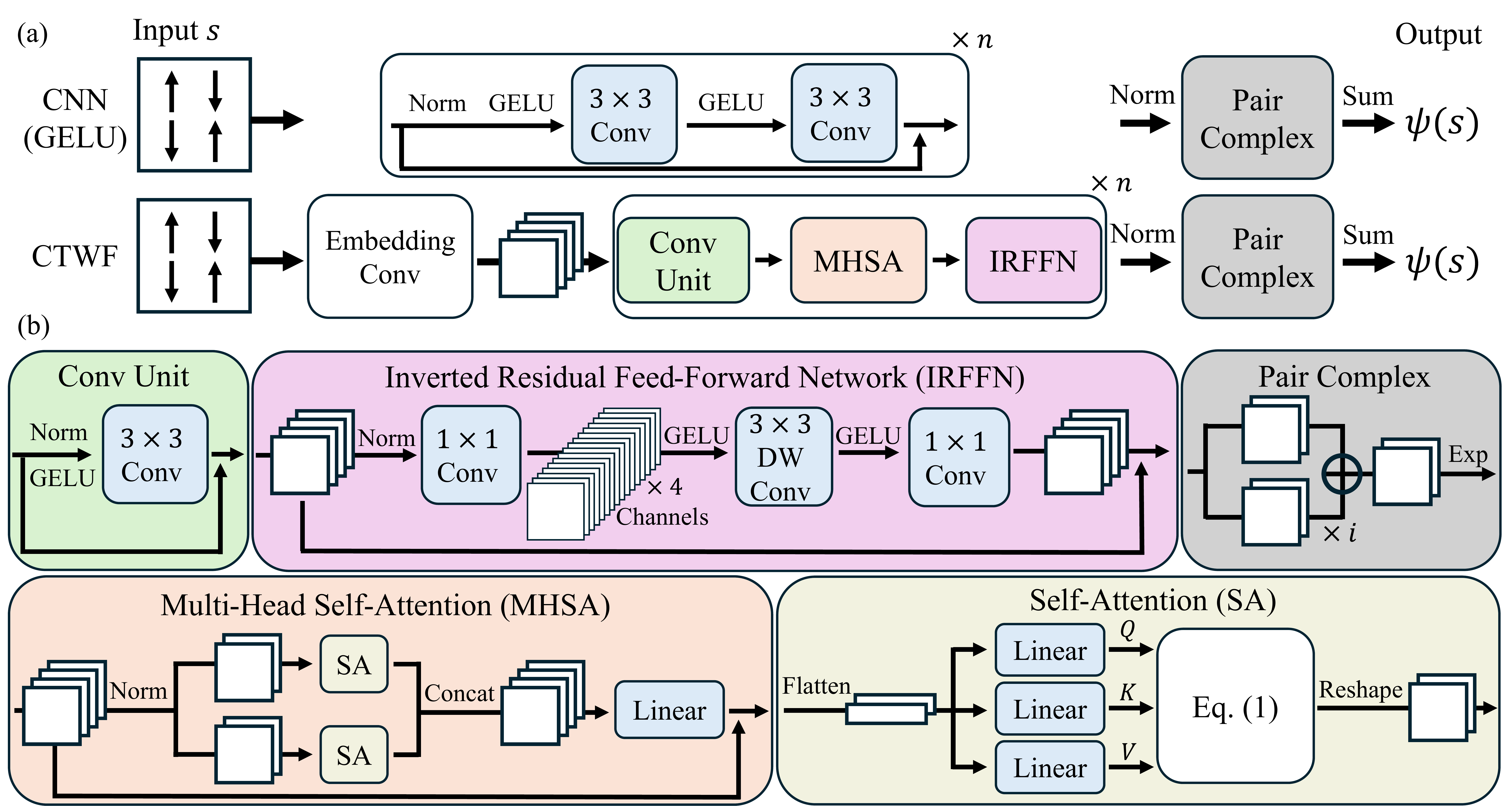}
    \caption{\textbf{(a)} Network architecture of the CNN (GELU) improved from Ref.\,\cite{Chen_NP24_MinSR} and the convolutional transformer wave function (CTWF).
    \textbf{(b)} Building blocks of CNN (GELU) and CTWF. The main block of CTWF consists of three consecutive parts, namely the convolutional unit (Conv Unit), the multi-head self-attention (MHSA), and the inverted residual feed-forward network (IRFFN). A normalization step (Norm) is applied at the beginning of all these parts by dividing the inputs by the expected initial standard deviation. The MHSA contains multiple self-attention (SA) blocks, each processed by Eq.\,\eqref{eq:attention}. The IRFFN consists of 3 consecutive convolutional layers, and the middle one with an expanded channel dimension is depthwise (DW) to reduce the computing cost. Finally, the pair complex activation function is applied to convert real values to complex outputs~\cite{Roth_PRB23_GCNN}. The illustration shown here has channel dimension $c=4$, self-attention token dimension $d=2$, and the number of heads $h=2$.}
    \label{fig:network}
\end{figure*}

In this work, we introduce the convolutional transformer wave function (CTWF) inspired by recently developed variants of transformers in computer vision tasks~\cite{Wu_ICCV21_CvT, Guo_CVPR22_CmT}.
In particular, we develop design choices motivated by physics principles and apply them to quantum spin systems.
The performance of CTWF for ground-state search is compared with the best previous results in the literature as well as a CNN (GELU) architecture, which we introduce here as an improvement from our previous work~\cite{Chen_NP24_MinSR}.
As a first challenging benchmark we consider the ground state search of the prototypical $10\times10$ $J_1$-$J_2$ Heisenberg model realizing a frustrated quantum magnet.
We find that both CTWF and CNN (GELU) outperform the best previous result given a similar amount of parameters in the underlying ANN.
Furthermore, we also benchmark the performance of CNN (GELU) and CTWF for quantum quench dynamics in the two-dimensional quantum Ising model.
We observe that the CTWF and CNN (GELU) provide reliable evolution trajectories for a much longer time as compared to the best existing result in the literature.
Overall, for the considered physics problems and design choices we conclude that both the CTWF and CNN (GELU) achieve superior performance as compared to existing literature.

\textit{Transformer architectures.}---
In the following, we revisit previously utilized transformer wave functions and introduce our CTWF whose structure is illustrated in Fig.\,\ref{fig:network}.
The core ingredient of transformer neural networks is the multi-head self-attention (MHSA), which can be viewed as a parallelization of $h$ attention heads~\cite{Vaswani_NIPS17_attention}. Each attention head takes the same input $x \in \mathbb{R}^{n \times c}$, where $n = l_x \cdot l_y$ represents the flattened spatial dimension and $c$ represents the embedded token dimension (or channel dimension in the context of CNN). The attention output is $A \in \mathbb{R}^{n \times d}$ given by 
$A_i = \sum_j \alpha(Q_i, K_j, P_{ij}) V_j$,
where $\alpha \in \mathbb{R}^{n \times n}$ is the attention coefficient, and $Q \in \mathbb{R}^{n \times d}$, $K \in \mathbb{R}^{n \times d}$, $V \in \mathbb{R}^{n \times d}$, and $P \in \mathbb{R}^{n \times n}$ are query, key, value, and positional encoding, respectively. Here $d$ is the so-called token dimension chosen as $d = c / h$ in this work. Below, we list multiple design choices for the attention mechanism in the existing literature.

\begin{table*}[t]
\caption{Performance comparison of various NQS architectures in the Heisenberg $J_1$-$J_2$ model on the $6\times6$ square lattice. The tested networks include complex-valued RBM with translation symmetry, CNN (GELU), the transformer with factored attention~\cite{Viteritti_PRL23_ViT, Viteritti_arxiv23_TransformerQSL, Rende_arxiv24_QK=0}, and the introduced CTWF. The number of real parameters $N_p$ and the number of multiply-accumulate operations (MACs) are also shown to indicate the complexity of architecture, which we attempt to keep at the same level for different networks.}
\begin{tabular}{C{2cm} | C{1.5cm} C{1.5cm} C{1.5cm} C{1cm} C{1cm} C{1cm} C{1cm} C{1cm} | C{1.5cm} C{1.5cm} C{1.5cm}}
NQS & $Q/K/V$ & IRFFN & RPE & $c$ & $d$ & $h$ & $N_p$ & MACs & $\epsilon_{\mathrm{rel}}$ & $\sigma^2/N$ & $I$ \\
\hline\hline
RBM &&&& 128 &&& 9472 & 331776 & 0.0153 & 0.0228 & 0.1376 \\
\hline
CNN (GELU) &&&& 16 &&& 7120 & 254016 & 0.0024 & \textbf{0.0047} & 0.0227 \\
\hline
\multirow{6}{*}{Transformer} & factored & linear & $\sqrt{}$ & 20 & 10 & 2 & 7992 & 317920 & 0.0040 & 0.0083 & 0.0283 \\
\cline{2-12}
 & linear & linear & $\sqrt{}$ & 18 & 9 & 2 & 7164 & 286092 & 0.0033 & 0.0066 & 0.0259\\
 & conv & linear & $\sqrt{}$ & 18 & 9 & 2 & 8136 & 321084 & 0.0031 & 0.0062 & 0.0236 \\
 & linear & conv & $\sqrt{}$ & 18 & 9 & 2 & 7884 & 309420 & \textbf{0.0023} & 0.0055 & \textbf{0.0137} \\
 & conv & conv & $\sqrt{}$ & 16 & 8 & 2 & 7208 & 283072 & 0.0025 & 0.0058 & 0.0186 \\
 & linear & conv & $\times$ & 16 & 8 & 2 & 7812 & 309420 & \textbf{0.0023} & 0.0053 & 0.0144\\
\end{tabular}
\label{table:6x6}
\end{table*}

\begin{itemize}
    \item In the original transformer~\cite{Vaswani_NIPS17_attention} and the vision transformer (ViT)~\cite{Dosovitskiy_ICLR21_ViT}, $Q = x W^Q$, $K = x W^K$, and $V = x W^V$, where $W^{Q/K/V} \in \mathbb{R}^{c \times d}$ are trainable parameters. $\alpha = \mathrm{softmax}(QK^T/\sqrt{d})$ with $\mathrm{softmax}(X)_{ij} = \exp (X_{ij}) / \sum_{k=1}^N \exp (X_{ik})$. The full attention formula is $A = \mathrm{softmax} (QK^T/\sqrt{d}) V$,
    which is most popular in machine learning, recently also adopted in quantum lattice models with translation symmetry broken by positional encoding~\cite{Cao_arxiv24_ViTimpurity}. The central issue in this design choice is how to efficiently utilize the spatial information of the input while keeping translation symmetry.

    \item In Ref.~\cite{Viteritti_PRL23_ViT, Rende_arxiv24_QK=0}, the simplified ViT with factored attention is applied to quantum lattice models. In the factored attention, $\alpha$ is directly given by the relative positional encoding (RPE)~\cite{Wu_ICCV21_RPE} $P_{ij} = p_{x_i-x_j, y_i-y_j}$, where $p \in \mathbb{R}^{l_x \times l_y}$ is trainable. The value $V$ is still $V = x W^V$, so the full attention is $A = P x W^V$.
    In this attention mechanism, $P$ is a linear transformation with a circular structure due to RPE, equivalent to a convolutional layer with full-size convolution kernel $p$. $W^V$ can be viewed as a linear layer or a $1\times1$ convolution. Therefore, the factored attention represents essentially two consecutive convolutional layers instead of a normal attention layer, as also discussed in~\cite{Rende_CP24_SRt}. This factored attention has been adopted in several quantum many-body problems~\cite{Viteritti_arxiv23_TransformerQSL, Roca-Jerat_arxiv24_TransformerLongrange, Herraiz-Lopez_arxiv24_LongrangeIsing}.

    \item The autoregressive transformer quantum state is implemented by masking out the contribution from $i > j$~\cite{Zhang_PRB23_MaskTransformer, Sprague_CP24_PatchedTransformer, Lange_arxiv24_Hybrid}. The attention is given by $A = \mathrm{softmax} (QK^T/\sqrt{d} + M) V$,
    where $M_{ij} = 0$ for $i \leq j$ and $M_{ij} = -\infty$ for $i > j$. This attention mechanism allows uncorrelated sampling due to its autoregressive property but hence suffers from the strong constraint on architecture and the broken translation symmetry.

\end{itemize}

In summary, while previous works have already considered transformers, the power of transformer architectures in NQS for quantum many-body systems has not been fully harnessed in existing literature.
This motivates us to introduce the aforementioned CTWF based on the recent progress of transformers in computer vision~\cite{Wu_ICCV21_CvT, Guo_CVPR22_CmT}. The attention is designed by adding RPE to the original attention mechanism, i.e.
\begin{equation} \label{eq:attention}
    A = \mathrm{softmax} \left( \frac{QK^T + P}{\sqrt{d}} \right) V,
\end{equation}
which implements an attention layer with translation symmetry. The attention outputs from different self-attention heads are concatenated after Eq.\,\eqref{eq:attention}.

The whole network is designed according to the convolutional transformer architecture in computer vision~\cite{Guo_CVPR22_CmT} with some modifications. The attention layer is sandwiched by two convolutional blocks, namely a convolutional unit and an inverted residual feed-forward network (IRFFN), to enhance its ability in representing local features and keep the translation symmetry across the whole network. We expect that the combination of MHSA and convolutional layers can help the network to encode both long-range and short-range correlations. An illustration of our network architecture is shown in Fig.\,\ref{fig:network}.

In order to challenge the performance of the CTWF architecture we compare also to CNN (GELU) as the most advanced CNN architecture to date, which we also introduce here as an improvement to the CNN in our previous work~\cite{Chen_NP24_MinSR}. The improvement mainly comes from the utilization of GELU activation~\cite{Hendrycks_arxiv23_GELU} instead of the previously chosen ReLU, which allows higher accuracy in ground-state search and stable evolution in dynamics.

\textit{Numerical results.}---
For the assessment of the performance of different network architectures in NQS ground state search, we train them for the paradigmatic frustrated $J_1$-$J_2$ Heisenberg model on a square lattice:
\begin{equation}
    \Ha = J_1 \sum_{\left<i,j\right>} {\bf{S}}_i \cdot {\bf{S}}_j 
    + J_2 \sum_{\left<\left<i,j\right>\right>} {\bf{S}}_i \cdot {\bf{S}}_j,
\end{equation}
where ${\bf{S}}_i = (S_i^x, S_i^y, S_i^z$) denotes spin-1/2 operators at site $i$. $\left<i,j\right>$ and $\left<\left<i,j\right>\right>$ indicate pairs of nearest and next-nearest neighbor sites, respectively. In this work, we specifically focus on the maximally frustrated point at $J_2/J_1 = 0.5$.

We assess the accuracy of the variational state $\ket{\psi}$ with respect to the exact ground state $\ket{\phi}$ through several quantities including the relative error of variational energy, a rescaled energy variance, and the infidelity. The relative error of energy is given by $\epsilon_{\mathrm{rel}} = (E - E_0)/|E_0|$
with $E = \braket{\psi|\Ha|\psi} / \braket{\psi|\psi}$ and $E_0 = \braket{\phi|\Ha|\phi} / \braket{\phi|\phi}$. The rescaled energy variance is defined as $\sigma^2 / N = (\braket{\Ha^2} - E^2) / N$,
where $N$ denotes the system size. Finally, the infidelity is $I = 1 - \braket{\psi|\phi} \braket{\phi|\psi} / \braket{\psi|\psi} \braket{\phi|\phi}$,
which is only available in small systems solvable by exact diagonalization (ED). $\epsilon_{\mathrm{rel}}$, $V$ and $I$ all tend to zero when the variational state $\ket{\psi}$ approaches the exact ground state $\ket{\phi}$.

We start by comparing different NQSs on the $6\times6$ lattice in order to identify an optimal network structure utilized later also for the larger system sizes. Each network is trained by stochastic reconfiguration (SR)~\cite{Sorella_PRL98_SR} with $10^4$ Monte-Carlo samples for $10^4$ steps. To ensure a fair comparison among different networks, we keep the number of network parameters $N_p$ at a similar level to control the network size, and the number of multiply-accumulate operations (MACs) at a similar level to control the runtime. In this way, the comparison targets not only the accuracy of the ground-state solution but rather the performance under similar usage of computational resources.

The performance of RBM, CNN (GELU), and different transformers are shown in Table\,\ref{table:6x6}. All deep networks studied in this work significantly outperform the shallow RBM, showing the necessity of modern deep NQSs for complex quantum systems. The factored attention, as we explained, is equivalent to a CNN with full kernels. As one can see its performance does not reach up to other transformers or CNNs with small kernels which appears superior as also observed in recent works~\cite{Chen_NP24_MinSR, Roth_PRB23_GCNN, Liang_MLST23_CNNJ1J2}.
We also compare different design choices for our transformer wave functions, as we will explain in the following. Firstly, the $Q$, $K$, and $V$ in Eq.\,\eqref{eq:attention} can be implemented by a convolutional layer~\cite{Wu_ICCV21_CvT} or a linear layer~\cite{Guo_CVPR22_CmT}. Secondly, the IRFFN block can be composed of linear or convolutional layers. Finally, the relative positional encoding (RPE) may or may not be present in Eq.\,\eqref{eq:attention}. In these tests, the combination of linear $Q/K/V$, convolutional IRFFN, and present PRE exhibits the best accuracy, which is the architecture shown in Fig.\,\ref{fig:network}. The optimal CTWF design choice achieves a similar level of accuracy compared with CNN (GELU). Some other design choices not shown in Table\,\ref{table:6x6} have also been tested, including replacing the normalization step by LayerNorm, utilizing ReLU instead of GELU, removing IRFFN, removing both convolutional unit and IRFFN, or employing other final activation functions, but we find that these variants do not improve the accuracy. These numerical experiments finally allow us to identify an optimal CTWF architecture displayed in Fig.\,~\ref{fig:network}, which will be now fixed for the following simulations.

\begin{figure}[t]
    \centering
    \vspace{-0.3cm}
    \hspace{-0.5cm}
    \subfigure{
        \includegraphics[height=6.6cm]{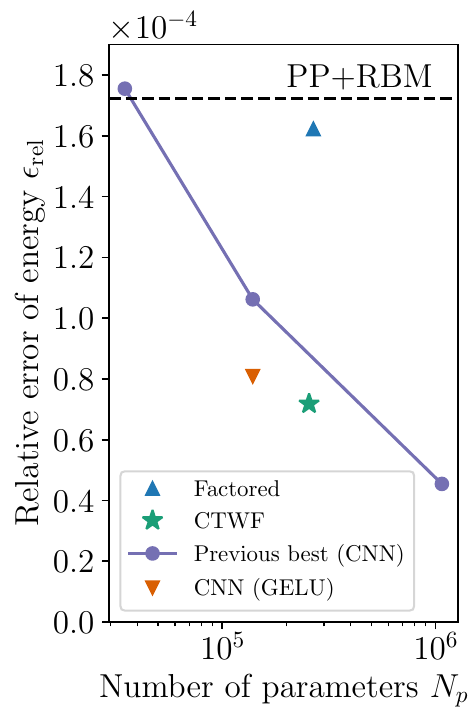}
    }
    \hspace{-0.5cm}
    \subfigure{
        \includegraphics[height=6.6cm]{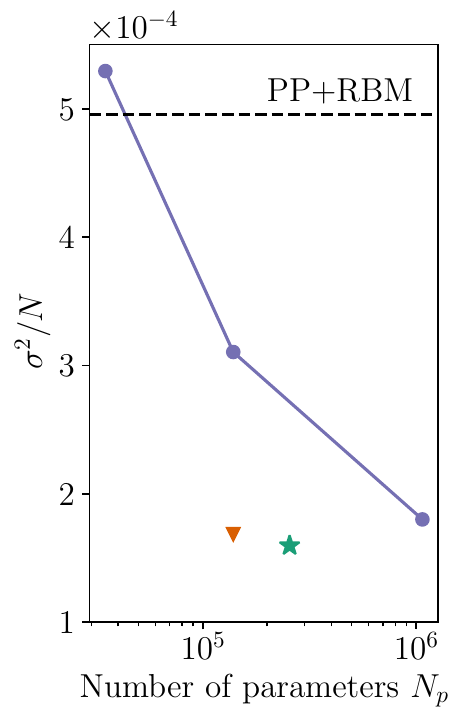}
    }
    \caption{The relative error of energy $\epsilon_\mathrm{rel}$ and the rescaled energy variance $\sigma^2/N$ in the $10\times10$ $J_1$-$J_2$ Heisenberg model at $J_2/J_1=0.5$. The results presented here include the combination of the pair product state (PP) and RBM~\cite{Nomura_PRX21_PPRBMJ1J2}, the best previous results given by a CNN in Ref.\,\cite{Chen_NP24_MinSR}, the transformer with factored attention~\cite{Rende_CP24_SRt}, the CNN (GELU), and the CTWF introduced in this work. The reference ground state energy is estimated by zero-variance extrapolation in Ref.\,\cite{Chen_NP24_MinSR}.}
    \label{fig:10x10J1J2}
\end{figure}

As a next step we now challenge the performance of the CTWF for the $10 \times 10$ $J_1$-$J_2$ Heisenberg model at $J_2 / J_1 = 0.5$ in Fig.\,\ref{fig:10x10J1J2}, choosing $n=5$, $c=48$, $d=12$, $h=4$, and $N_p = 255440$.
The result of the CNN (GELU) with $n=8$ and $c=32$ is also included. Apart from the inherent translation symmetry in these networks, we also apply symmetry projections including spatial reflection, rotation, and spin-flip, which amounts to 16 symmetry group elements in total. Here, the optimization is performed with $10^4$ Monte-Carlo samples and MinSR~\cite{Chen_NP24_MinSR}.

Given a similar amount of parameters, the variational energy of our CTWF and CNN (GELU) significantly outperforms the factored attention~\cite{Rende_CP24_SRt}. With more samples $N_s=2^{14}$ and more parameters $N_p = 434760$, the factored attention is possible to reach variational energy $-0.4976764(7)$ and $\epsilon_{\mathrm{rel}}=7.9 \times 10^{-5}$~\cite{communications} compatible with the best previous CNN result, while still less accurate than CTWF. The result of autoregressive transformers is not shown due to the lack of reference, but here we present the result from an RNN~\cite{Hibat-Allah_PRR20_NQSRNN} for illustration. The $\epsilon_\mathrm{rel}$ and rescaled energy variance $\sigma^2/N$ of this RNN are respectively $4.7 \times 10^{-3}$ and $1.0 \times 10^{-3}$~\cite{Wu_Science24_Vscore}, which are significantly higher than the values in Fig.\,\ref{fig:10x10J1J2} and are hence not included in the figure. This suggests that the constrained network architecture and broken translation symmetry might be the key limiting factors in autoregressive NQSs including autoregressive transformers. Compared with the best previous result~\cite{Chen_NP24_MinSR}, CTWF and CNN (GELU) show similar accuracy in variational energy but produce a lower variance. These results show that these architectures lead to competitive variational wave functions for frustrated quantum magnets, which can serve as alternatives to existing CNN and transformer wave functions in future applications.

\begin{figure}[t]
    \centering
    \includegraphics[width=1.0\linewidth]{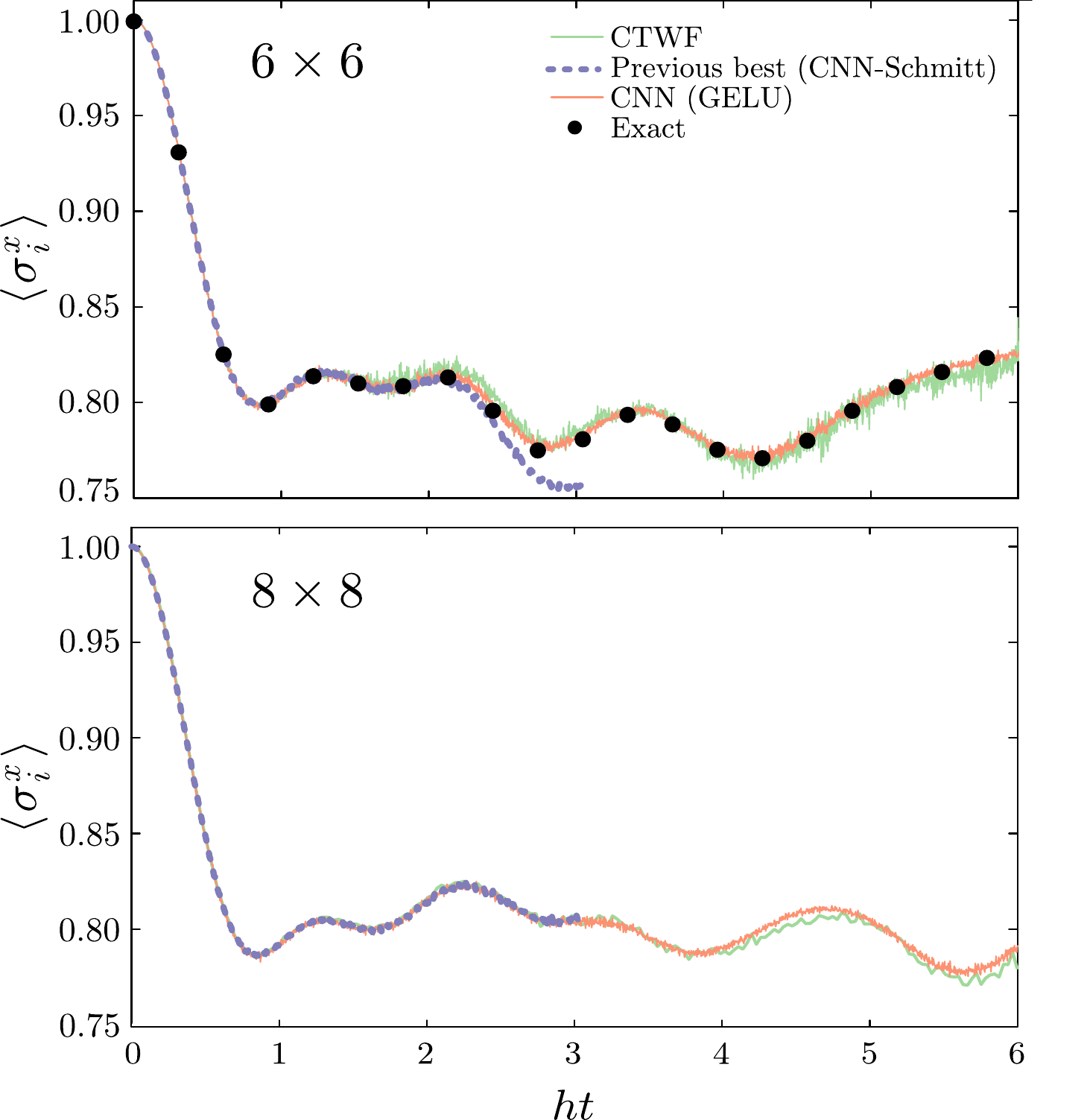}
    \caption{Quench dynamics in the $6 \times 6$ and $8 \times 8$ transverse-field Ising model at the critical point simulated by CTWF. The best previous TDVP result (CNN-Schmitt)~\cite{Schmitt_PRL20_NQSdynamics} is shown for reference.}
    \label{fig:dynamics}
\end{figure}

Furthermore, we emphasize that the application of CTWF is not limited to ground-state searches. As an example, we present a simulation of real-time dynamics in quantum many-body systems using CTWF.  
For this study, we employ the prototypical transverse-field Ising model,  
\begin{equation} \label{eq:TFIsing}
    \mathcal{H} = -J \sum_{\braket{i,j}} \sigma^z_i \sigma^z_j - h \sum_i \sigma^x_i,
\end{equation}  
which has been widely used in various neural network quantum dynamics studies~\cite{Schmitt_PRL20_NQSdynamics, Schmitt_SA22_QPTdynamics, Sinibaldi_Q23_UnbiaseTDVP}.   
The simulation is conducted at the most challenging parameter point of the underlying quantum phase transition, \( h/J = 3.04438(2) \)~\cite{Blote_PRE02_TFIsing}, on \(6 \times 6\) and \( 8 \times 8 \) square lattices.  
The quench dynamics begins from a paramagnetically polarized state, \( \ket{\psi_0} = \ket{\rightarrow}^{\otimes N} \), and evolves under the Hamiltonian in Eq.\,\eqref{eq:TFIsing}.  
The real-time evolution is performed using the time-dependent variational principle (TDVP) with a fixed time interval of \( J\tau = 10^{-3} \), employing the second-order Heun method.
To further improve precision, symmetry projections—including spatial reflection, rotation, and spin-flip—are applied during the simulation.

We evaluate performance by comparing the expectation value \( \langle \sigma_i^x(t) \rangle \) obtained using the CNN (GELU), the CTWF, and the best previous TDVP result (denoted as  CNN-Schmitt) in Ref.\,\cite{Schmitt_PRL20_NQSdynamics}; see Fig.\,\ref{fig:dynamics}.
For \(6 \times 6\), we also compare this with the exact result obtained by integrating the time-dependent Schrodinger equation.
For both system sizes, our approach successfully extends the stability of the dynamics for a longer time using both CNN (GELU) and CTWF compared to previous benchmarks.
This highlights not only the improved accuracy and robustness of CTWF and CNN (GELU) but also demonstrates that CTWF is a powerful neural quantum state for studying quantum dynamics, in addition to its well-established effectiveness in ground-state calculations.

\textit{Discussion.}---
In this work, we have introduced the convolution transformer wave function (CTWF).
We have found compelling evidence that this NQS architecture exhibits outstanding performance for both ground-state search and non-equilibrium quantum dynamics as compared to existing results in the literature.
While these results highlight the potential of transformers for solving quantum lattice models, it is also important to emphasize that no definite conclusion on the superiority of transformers can be reached at this point, as an also introduced CNN (GELU) network structure yields comparable results.
However, considering that the study of transformer wave functions is still at a comparatively early stage, it appears possible that CTWF might outperform CNNs upon further developments.

One issue we face in this work is the computing cost of transformers due to self-attention, which originates from the $\mathcal{O}(N^2)$ complexity of self-attention as compared to the $\mathcal{O}(N)$ complexity of CNN, where $N$ is the system size. With a suitable choice of the embedding stride, nevertheless, the system size can be reduced to $\sqrt{N}$ to keep $\mathcal{O}(N)$ complexity. As transformers have been popular in the community of artificial intelligence, we also expect the relevant progress in theory, software, and hardware can benefit our CTWF in the future and help us to scale CTWF to more parameters and larger systems. We therefore expect great potential for further improvements.

Furthermore, the CNN architecture is designed for encoding local features, whereas the self-attention in transformers can potentially more efficiently express long-range correlations. Therefore, we expect self-attention to be an important structure for ground states of Hamiltonians with long-range interactions or quantum dynamics in large systems. In these cases, the CTWF with both CNN and self-attention might be a good balance for expressing efficiently both local and global correlations.

The data of Fig.\,\ref{fig:10x10J1J2} and Fig.\,\ref{fig:dynamics} is presented in the Zenodo repository~\cite{Zenodo_data}.

\begin{acknowledgments}
The exact results including ground state and real-time evolution are computed by the QuSpin package~\cite{Weinberg_SPP17_QuSpinI}. The NQS dynamics is simulated with the jVMC package~\cite{Schmitt_SP22_jVMC}. We thank Markus Schmitt for providing the data for Fig.\,\ref{fig:dynamics}. We also acknowledge fruitful discussions with C. Roth, L. L. Viteritti, R. Rende, and F. Becca. This project has received funding from the European Research Council (ERC) under the European Union’s Horizon 2020 research and innovation programme (grant agreement No. 853443). This work was supported by the German Research Foundation DFG via project 492547816 (TRR 360). We gratefully acknowledge the scientific support and high-performance computing resources provided by the Scientific Computing Core at the Flatiron Institute, the LiCCA HPC cluster of the University of Augsburg, and the Erlangen National High Performance Computing Center (NHR) of the Friedrich-Alexander-Universität Erlangen-Nürnberg (NHR Project No. nqsQuMat). The LiCCA cluster is co-funded by the Deutsche Forschungsgemeinschaft (DFG, German Research Foundation) – Project-ID 499211671. NHR funding is provided by federal and Bavarian state authorities. NHR@FAU hardware is partially funded by the German Research Foundation (Grant No. 440719683). 
\end{acknowledgments}

\bibliography{reference}

\end{document}